\documentclass[aps,prl,preprint,tightenlines,superscriptaddress,showpacs,byrevtex]{revtex4}
\usepackage{graphicx} 
\usepackage{dcolumn}  

\graphicspath{{ps}}

\def\bhck {B \to h_c K}

\def\etacg {h_c \to  \eta_c \gamma}
\def\ekskpi {\eta_{c} \to K_S^0K^{-}\pi^{+}}
\def\epp {\eta_{c} \rightarrow p \bar{p}}
\def\kskpi {K_S^0K^{\pm}\pi^{\mp}}
\def\pp {p \bar{p}}
\def\mb {M_{\rm bc}}
\def\de {\Delta E}

\begin{document}


\preprint{\vbox{ \hbox{Belle Preprint 2006-15   }
                 \hbox{KEK Preprint 2006-15}
}}

\title{ \quad\\[0.5cm]  Search for the $h_c$ meson in $B^{\pm}\to h_c K^{\pm}$}


\affiliation{Budker Institute of Nuclear Physics, Novosibirsk}
\affiliation{Chiba University, Chiba}
\affiliation{Chonnam National University, Kwangju}
\affiliation{University of Cincinnati, Cincinnati, Ohio 45221}
\affiliation{University of Hawaii, Honolulu, Hawaii 96822}
\affiliation{High Energy Accelerator Research Organization (KEK), Tsukuba}
\affiliation{University of Illinois at Urbana-Champaign, Urbana, Illinois 61801}
\affiliation{Institute of High Energy Physics, Chinese Academy of Sciences, Beijing}
\affiliation{Institute of High Energy Physics, Vienna}
\affiliation{Institute of High Energy Physics, Protvino}
\affiliation{Institute for Theoretical and Experimental Physics, Moscow}
\affiliation{J. Stefan Institute, Ljubljana}
\affiliation{Kanagawa University, Yokohama}
\affiliation{Korea University, Seoul}
\affiliation{Kyungpook National University, Taegu}
\affiliation{Swiss Federal Institute of Technology of Lausanne, EPFL, Lausanne}
\affiliation{University of Ljubljana, Ljubljana}
\affiliation{University of Maribor, Maribor}
\affiliation{University of Melbourne, Victoria}
\affiliation{Nagoya University, Nagoya}
\affiliation{Nara Women's University, Nara}
\affiliation{National Central University, Chung-li}
\affiliation{National United University, Miao Li}
\affiliation{Department of Physics, National Taiwan University, Taipei}
\affiliation{H. Niewodniczanski Institute of Nuclear Physics, Krakow}
\affiliation{Niigata University, Niigata}
\affiliation{Nova Gorica Polytechnic, Nova Gorica}
\affiliation{Osaka City University, Osaka}
\affiliation{Osaka University, Osaka}
\affiliation{Panjab University, Chandigarh}
\affiliation{Peking University, Beijing}
\affiliation{Princeton University, Princeton, New Jersey 08544}
\affiliation{RIKEN BNL Research Center, Upton, New York 11973}
\affiliation{University of Science and Technology of China, Hefei}
\affiliation{Seoul National University, Seoul}
\affiliation{Sungkyunkwan University, Suwon}
\affiliation{University of Sydney, Sydney NSW}
\affiliation{Tata Institute of Fundamental Research, Bombay}
\affiliation{Toho University, Funabashi}
\affiliation{Tohoku Gakuin University, Tagajo}
\affiliation{Tohoku University, Sendai}
\affiliation{Department of Physics, University of Tokyo, Tokyo}
\affiliation{Tokyo Institute of Technology, Tokyo}
\affiliation{Tokyo Metropolitan University, Tokyo}
\affiliation{Tokyo University of Agriculture and Technology, Tokyo}
\affiliation{Virginia Polytechnic Institute and State University, Blacksburg, Virginia 24061}
\affiliation{Yonsei University, Seoul}
   \author{F.~Fang}\affiliation{University of Hawaii, Honolulu,
Hawaii 96822} 
   \author{T.~E.~Browder}\affiliation{University of Hawaii, Honolulu, Hawaii 96822} 
   \author{K.~Abe}\affiliation{High Energy Accelerator Research Organization (KEK), Tsukuba} 
   \author{K.~Abe}\affiliation{Tohoku Gakuin University, Tagajo} 
   \author{I.~Adachi}\affiliation{High Energy Accelerator Research Organization (KEK), Tsukuba} 
   \author{H.~Aihara}\affiliation{Department of Physics, University of Tokyo, Tokyo} 
   \author{D.~Anipko}\affiliation{Budker Institute of Nuclear Physics, Novosibirsk} 
   \author{T.~Aushev}\affiliation{Institute for Theoretical and Experimental Physics, Moscow} 
   \author{A.~M.~Bakich}\affiliation{University of Sydney, Sydney NSW} 
   \author{V.~Balagura}\affiliation{Institute for Theoretical and Experimental Physics, Moscow} 
% ...
\collaboration{The Belle Collaboration}


\noaffiliation

\begin{abstract}

We report a search for the $h_c$ meson via the decay chain $B^{\pm}\to
h_c K^{\pm}$, $\etacg$ with $\eta_c \to K_S^0K^{\pm}\pi^{\mp} $ and $p\bar{p}$. No
significant signals are observed. We obtain upper limits on the
branching fractions for $B^{\pm} \to \eta_c \gamma K^{\pm}$ in bins of the
$\eta_c\gamma$ invariant mass. The results are based on an analysis of
253 fb$^{-1}$ of data collected by the Belle detector at the KEKB $e^+
e^-$ collider.

\end{abstract}


\maketitle

\tighten

{\renewcommand{\thefootnote}{\fnsymbol{footnote}}}
\setcounter{footnote}{0}

The $h_c$ meson is the $1^1P_1$ spin singlet state of $c\bar{c}$,
which is one of the missing states in the charmonium spectrum below
$D\bar{D}$ threshold. It is expected to be a narrow resonance 
($\Gamma_{h_c}<1.1 $ MeV/$c^2$) that lies between $J/\psi(1S)$ and
$\psi(2S)$. The predicted masses of $h_c$ vary and are summarized in
Ref.~\cite{ref:mhc}. A typical value is less than 10 MeV from
the center of gravity of the $1^3P_J$ states ($\chi_{c0}$, $\chi_{c1}$
and $\chi_{c2}$), which is
$M_{c.o.g.}=(M_{\chi_{c0}}+3M_{\chi_{c1}}+5M_{\chi_{c2}})/9=3525.4\pm0.1$
MeV/$c^2$. The $h_c$ meson should decay dominantly to $\eta_c\gamma$ with a
branching fraction of about
$50\%$~\cite{ref:mhc,ref:hcdecay,ref:suzuki}. 

The E760 collaboration has reported an enhancement in the $p\bar{p}\to
h_c\to J/\psi \pi^0$ cross section and identified it as the $1^1P_1$
state with a mass of $3526.2\pm 0.25$ MeV/$c^2$~\cite{ref:e760}. This
result was not confirmed by the subsequent experiment E835 with
significantly higher statistics. However, E835\cite{ref:e835} 
reported promising evidence for the $h_c$ in $h_c\to \eta_c\gamma$. 
Recently, CLEO\cite{ref:cleo} has reported the observation of 
$h_c\to \eta_c\gamma$ at a mass of $M= 3524.4\pm 0.6 \pm 0.4$ MeV/$c^2$. 
The masses obtained by CLEO and E835 are within 1 MeV of $M_{c.o.g.}$. 


M. Suzuki~\cite{ref:suzuki} and others~\cite{ref:others} have
 proposed using the decay chain
$B\to h_c K$, $h_c \to  \eta_c \gamma$ to look for the $h_c$ meson.
Other charmonium candidates including the $\eta_c(2S)$\cite{ref:etac_2s},
$X(3872)$\cite{ref:x3872} and $Y(3940)$\cite{ref:y3940} were first
 observed in two-body $B$ 
decays, where the kinematic constraints from the exclusive $B$ decay and 
production at threshold provide substantial background reduction.
The decay amplitudes for $\bhck$ and $B\to \chi_{c0,2} K$
vanish in the factorization limit. The branching fractions
for $B^{+} \to \chi_{c0,2} K^{+}$\cite{ref:cc} have been measured and the
results are given below~\cite{ref:chic0},~\cite{ref:chic2}: 
\begin{eqnarray}
{\mathcal{B}}(B^{+}\to\chi_{c0}K^{+})&=&(1.34\pm 0.45 \pm 0.15\pm
0.04)\times 10^{-4} ({\rm BaBar}),
\nonumber \\
{\mathcal{B}}(B^{+}\to\chi_{c0}K^{+})&=&(1.12\pm 0.12 \pm 0.18\pm
0.08)\times 10^{-4} ({\rm Belle}),
\nonumber \\
{\mathcal{B}}(B^{+}\to\chi_{c2}K^{+})& <&0.3\times10^{-4}.
\end{eqnarray}
The fairly large branching fraction for $B^{+} \to \chi_{c0} K^{+}$
suggests that non-factorizable contributions in $B$ decays to
charmonium can be sizable. The decay $\bhck$ may occur via the
color octet mechanism~\cite{ref:beneke} or re-scattering
processes~\cite{ref:colangelo} at a rate comparable to that of the
factorization allowed decay mode $B \to \chi_{c1}K$. Thus, measurement
of the branching fraction 
for $B\to h_c K$ will provide useful information on
non-factorizable contributions in $B$ to charmonium decays.

Here we present the results of a search for $B^+\to h_c K^+$, 
$h_c\to  \eta_c\gamma$ with $\eta_c \to K_S^0 K^{\pm}\pi^{\mp} $ and $p
\bar{p}$ using a 253 fb$^{-1}$ data sample, which contains
$275 \times 10^6$ produced $B\bar{B}$ pairs. The data were 
collected at the $\Upsilon(4S)$ resonance with the Belle detector at
the KEKB $e^+ e^-$ collider\cite{KEKB}. In addition, we use a 28
fb$^{-1}$ data sample collected at an energy 60 MeV below resonance to
measure the continuum background.


The Belle detector is a large-solid-angle magnetic
spectrometer that consists of a silicon vertex detector (SVD),
a 50-layer central drift chamber (CDC), an array of
aerogel threshold \v{C}erenkov counters (ACC), 
a barrel-like arrangement of time-of-flight
scintillation counters (TOF), and an electromagnetic calorimeter (ECL)
comprised of CsI(Tl) crystals located inside 
a super-conducting solenoid coil that provides a 1.5~T
magnetic field.  An iron flux-return located outside of
the coil is instrumented to detect $K_L^0$ mesons and to identify
muons (KLM).  The detector is described in detail elsewhere~\cite{Belle}.
A sample of 152 million $B\bar{B}$ pairs was taken with a 2.0 cm
radius beampipe and a 3-layer silicon vertex detector; another
sample of 123
million $B\bar{B}$ pairs was taken with a 1.5 cm radius beampipe, a
4-layer silicon detector and a small-cell inner drift chamber\cite{Ushiroda}.

Event selection criteria were determined using the figure of merit,
which is defined as $S/\sqrt{S+B}$, where $B$ is
the number of background events and $S$ is the number of signal
events in a GEANT-based Monte Carlo simulation. 
We assume ${\cal{B}}(B^+\to h_c K^+)=3\times 10^{-4}$ and
${\cal{B}}(h_c \to   \eta_c \gamma)=0.5$ for the signal and determine
the background $B$ from the sideband data. Signal events are simulated for five
different values of the $h_c$ mass, which are $M_{h_c}=3.23$,
$3.33$, $3.43$, $3.527$ and $3.63$
GeV/$c^2$, and assuming the intrinsic width $\Gamma_{h_c}=1$ MeV/$c^2$. 
We determine the final optimization of selection requirements with the
$M_{h_c}=3.527$ GeV/$c^2$ MC sample.

We select well measured charged tracks with impact parameters with
respect to the interaction point (IP) of less than 0.3 cm in the radial
direction and less than 5 cm in the $z$ direction, which is opposite to
the positron beam direction. The tracks are required to have the
transverse momenta greater than $50$ MeV/$c$ and have more than 6
axial and 2 stereo CDC hits.

Particle identification likelihoods for the pion and kaon particle
hypotheses are calculated by combining information from the TOF and
ACC systems with $dE/dx$ measurements in the CDC. To identify
kaons, we require the kaon likelihood ratio, $L_K/(L_K+L_\pi)$,
to be greater than 0.6, which is 89\% efficient for kaons with a 8\%
misidentification rate for pions. For the charged kaons that come
directly from the $B$ meson 
rather than from the subsequent decay of the $\eta_c$, the kaon
likelihood ratio is required to be greater than 0.5. In addition, we
remove all kaon candidates that are consistent with being either
protons or electrons. To identify pions, we require $L_K/(L_K+L_\pi)$
to be smaller than 0.7, which is 94\% efficient for pions with a 12\%
misidentification rate for kaons.

Protons and antiprotons are identified using all particle
identification systems and are required to have proton likelihood
ratios [$L_p/(L_p+L_K)$ and $L_p/(L_p+L_\pi)$] greater than
0.5. Proton candidates that are electron-like according to the
information from the ECL are vetoed. This selection
is 90\% efficient for protons with a 6\% misidentification rate for
kaons and a 3\% misidentification rate for pions.

We select $K^0_{S}\to \pi^+\pi^-$ candidates from pairs of oppositely
charged tracks that are consistent with the pion hypothesis to form
common vertices and lie within the mass window $0.482$ $\rm{GeV}/c^2$
$<$ $M(\pi^{+}\pi^{-})$ $<$ $0.514$ $\rm{GeV}/c^2$, which corresponds
to $\pm 4\sigma$. The $K^0_{S}$ vertex is required to be displaced
from the IP; the vertex direction from the IP is required to be
consistent with the $K^0_{S}$ flight direction. The $K^0_{S}$
requirements are described in detail elsewhere~\cite{ref:ks}.


We reconstruct $\eta_{c}$ candidates in the $K_S^0 K^{\pm}\pi^{\mp}$
and $p\bar{p}$ decay modes. The $\eta_{c}$ candidate is required to
have an invariant mass in the range between 2.935 and 3.035
GeV/$c^2$. In order to reduce the combinatorial background, the
charged daughters of the $\eta_{c}$ are required to come from a common
vertex that is consistent with the interaction point profile.

Photon candidates for the decay $h_c\to  \eta_c \gamma $
are selected from ECL clusters that are not associated with charged tracks
extrapolated from the CDC. We require the photons have energies above
60 MeV, and at least five crystal hits.

To isolate the signal, we form the beam constrained mass 
$M_{\rm bc}=\sqrt{ E_{\rm beam}^2 - \vec{P}_{\rm recon}^2 }$ and
energy difference $\Delta E  = E_{\rm recon} - E_{\rm beam}$, where
$E_{\rm beam}$, $E_{\rm recon}$ and $\vec{P}_{\rm recon}$ are the beam
energy, the reconstructed energy and the reconstructed momentum of a
$B$ candidate in the $\Upsilon(4S)$ center of mass frame. The signal
region for $M_{\rm bc}$ is $5.270<M_{\rm bc}<5.290$ GeV/$c^2$. The
signal region for $\Delta E$ is $-50<\Delta E<35$ MeV, which
corresponds to $\pm 2.5\sigma$ where $\sigma$ is the resolution
determined from a Gaussian fit to the Monte Carlo simulation. If more
than one signal candidate is found in an event, we select
the one with the largest invariant mass $M(K^+\gamma)$, where the kaon
comes from the $B$, the best $\chi^2$ of the $\eta_c$ vertex and the
invariant mass $M(\kskpi)$ or $M(\pp)$ closest to the nominal mass of
the $\eta_c$. These requirements are imposed in the order listed till
only one candidate is selected.

To suppress the large background from continuum $e^+ e^-\to q \bar{q}$
where $q=u, d, s, c$, we first remove events with the normalized second
Fox-Wolfram moment $R_2>0.5$. We then form a likelihood ratio using
two variables. Six modified Fox-Wolfram moments\cite{ref:belle_fw} and
the cosine of the thrust angle are combined into a Fisher discriminant
$\cal{F}$. For signal Monte Carlo and continuum data, we form probability
density functions for this Fisher discriminant, and the cosine of the
$B$ decay angle with respect to the $z$ axis ($\cos\theta_B$). 
We then calculate the likelihood 
ratio ${\cal R}$ $={\cal L}_S/({\cal L}_S+{\cal
L}_{BG})$ for the $B^+\to h_c K^+$ signal Monte Carlo and continuum data. The
likelihood ratio $\cal{R}$ for the $\ekskpi$ mode is required to be
greater than 0.7. This requirement retains 70\% of the signal while
removing 92\% of the continuum background. For the $p\bar{p}$ mode, which
has less continuum background, we require $\cal{R}$ to be greater
than 0.6.

In addition to backgrounds from continuum there are also backgrounds
from other $B$ decays.
To investigate these backgrounds, we use a sample of
$379\times 10^6$ $B\bar{B}$ Monte Carlo events. We find that the
dominant backgrounds come from $B^{+} \to \eta_c K^{*+}$,
$K^{*+} \to K^{+} \pi^0$ and $B^0 \to \eta_c K^{*0}$, $K^{*0} \to
K^+ \pi^-$. These backgrounds peak in the $M_{\rm bc}$
distributions. We reject the events if the photon combined with any
other photon makes a $\pi^0$ candidate with
$0.114<M(\gamma\gamma)<0.151$ MeV/$c^2$. This $\pi^0$ veto requirement is
83\% efficient for signal and removes 51\% of background $\pi^0$'s. 
We also require the cosine of the angle in the $\eta_c\gamma$ rest frame
between the $\gamma$ and the
kaon coming from the $B$ candidate be smaller than 0.6 (0.9) if the
invariant mass $M$($ \eta_c \gamma$) is smaller (greater) than 3.5
GeV. These requirements retain 70\% (85\%) of the signal while removing
67\% (72\%) of the $B \to \eta_c K^{*}$ backgrounds.

We also find backgrounds from $B^+ \to \eta_c K^+$ and $B\to
D_s^{(*)+}\bar{D}^{(*)}$, which peak in the $M_{\rm bc}$
distributions. We remove the $B^+ \to \eta_c K^+$ background if the
$\Delta E$ for this decay mode is between $-60$ and $+60$ MeV ($\pm
6\sigma$). This requirement is about $100\%$ efficient for the
signal. We also apply a $D_s^+$ veto if the invariant mass $M(K^+K_S)$
is in the range $1.938<M(K^+K_S)<1.998$ GeV/$c^2$ ($\pm3\sigma$). This
requirement is $94\%$ efficient for the signal.

After all the event selection requirements are applied, no significant
peaking $B\bar{B}$ backgrounds are observed. A component for the 
$B\bar{B}$ background is included in the fits to data. In addition,
uncertainties in the $B \bar{B}$ background composition are included
in the systematic error.

\begin{figure}[htb]
\begin{center}
\includegraphics[width=0.48\textwidth]{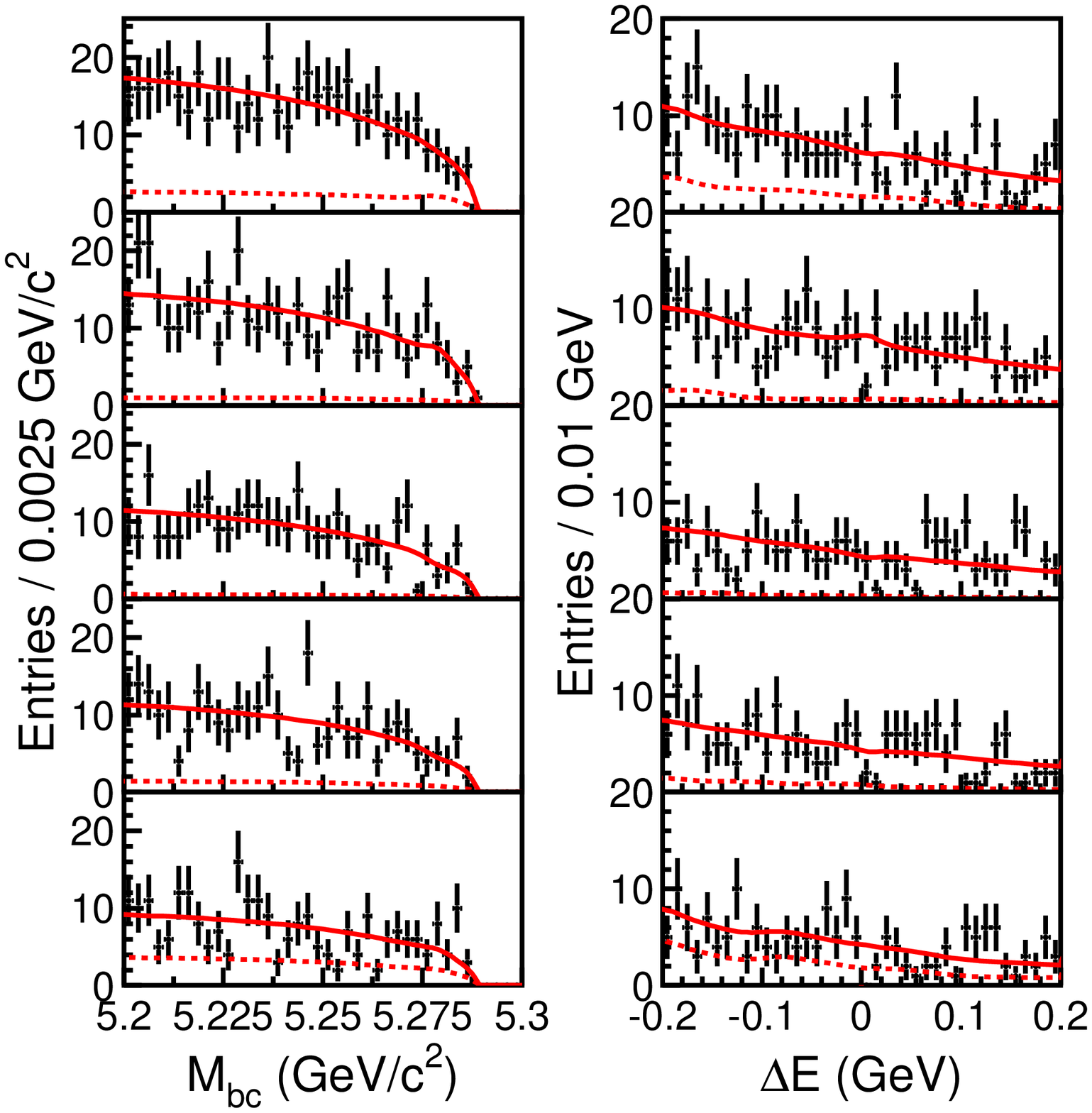}
\includegraphics[width=0.48\textwidth]{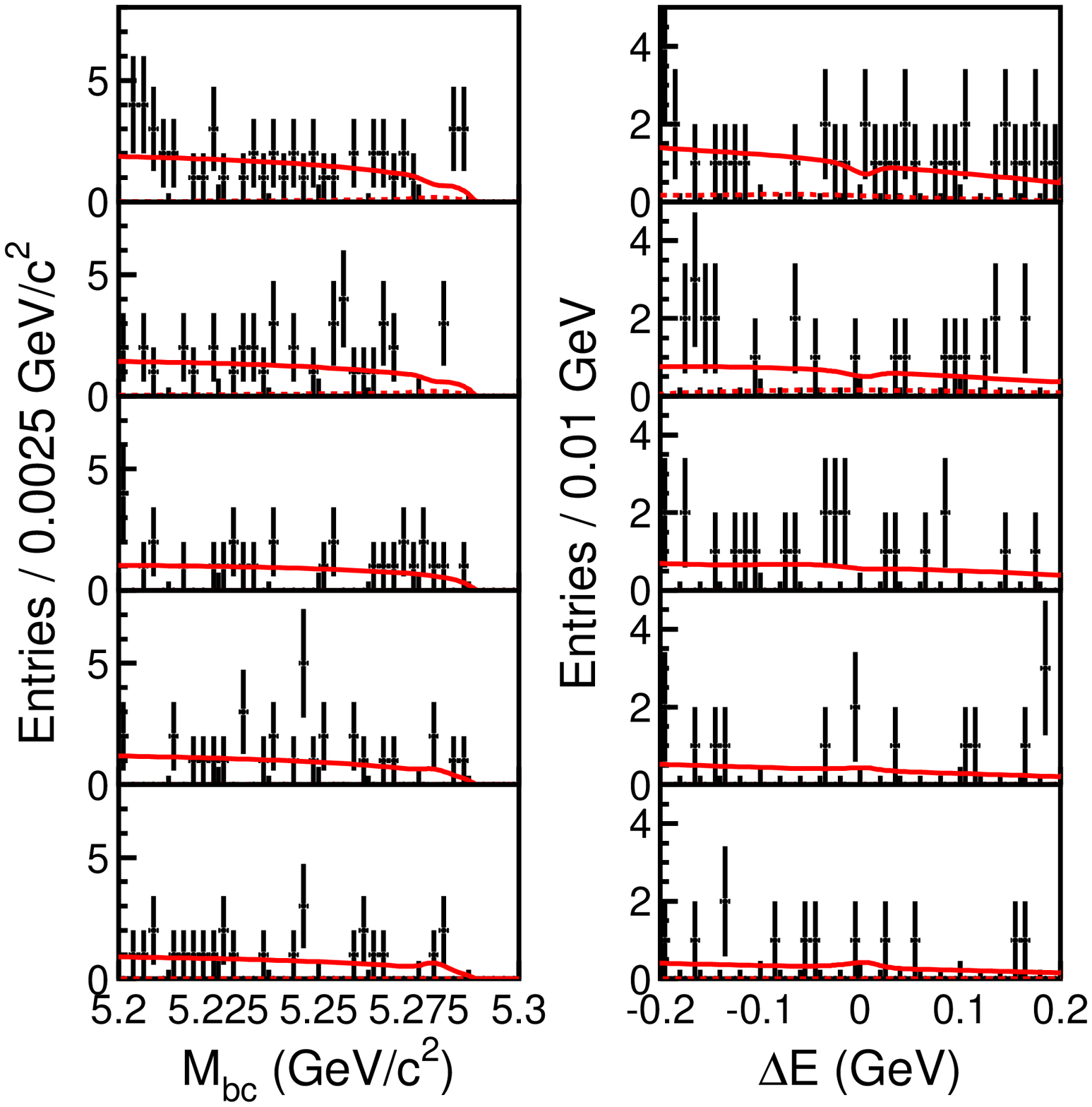}
\caption[]{The $\mb$ distributions in the $\de$ signal region and the
  $\de$ distributions in the $\mb$ signal region for the $\ekskpi$ (left) and
$\epp$ modes (right) in 100 MeV bins of $M( \eta_c\gamma)$ for $3.17 <
M( \eta_c\gamma) \le 3.67$ GeV/$c^2$. The distributions are shown in the  
increasing order of $M( \eta_c\gamma)$ from the top to the bottom. The
solid curves are the results of the fits. The dashed curves represent
background components from $B$ decays. }  
\label{fig:rangeI}
\end{center}
\end{figure}

In Fig.~\ref{fig:rangeI}, we show the $M_{\rm bc}$ and $\de$
distributions in five 100 MeV/$c^2$ bins of $M( \eta_c  \gamma)$,  which
correspond to a $\pm 3\sigma$ range around the central values used in
the signal MC. We determine signal yields from unbinned two-dimensional 
maximum-likelihood fits to the $\mb$-$\de$ distributions in
the in region $5.2<\mb<5.3$ GeV/$c^2$ and $-0.2<\de<0.2$ GeV. We use a
Gaussian function for $\mb$ and a double Gaussian for $\de$
to model the signal. The mean of the Gaussian for $\mb$ is fixed to
$5.279$ GeV/$c^2$, which is determined from a $B^+ \rightarrow
D^{*0}\pi^+$ data sample, and other parameters are fixed to the
values from MC simulation. 
The functions used to model the backgrounds for other $B$
decays are determined from MC. The continuum background is modeled with an
ARGUS background function that behaves like phase space near the
kinematic boundary~\cite{ref:argus91} for $\mb$ and a linear function
for $\de$. We determine the shape parameters for the background functions
from a fit to the off-resonance data sample.  We find that the
background shapes do not depend on $M(\eta_c\gamma)$.
Therefore, we combine the data in the range
$3.17<M( \eta_c \gamma)<3.72$ GeV/$c^2$ to increase statistics.

Because the mass of the $h_c$ is not well established, we fit the
$M_{\rm bc}$-$\de$ distributions in the five 100 MeV/$c^2$ bins of
$M( \eta_c \gamma)$. The results of the fits are shown in
Fig.~\ref{fig:rangeI}. The signal yields and the detection
efficiencies, which are determined from the signal MC samples
described above, are given in Table~\ref{tab:fit}. No significant
signals are observed for $3.17 < M( \eta_c\gamma) \le 3.67$
GeV/$c^2$.

~\begin{figure}[htb]
\begin{center}
\includegraphics[width=0.48\textwidth]{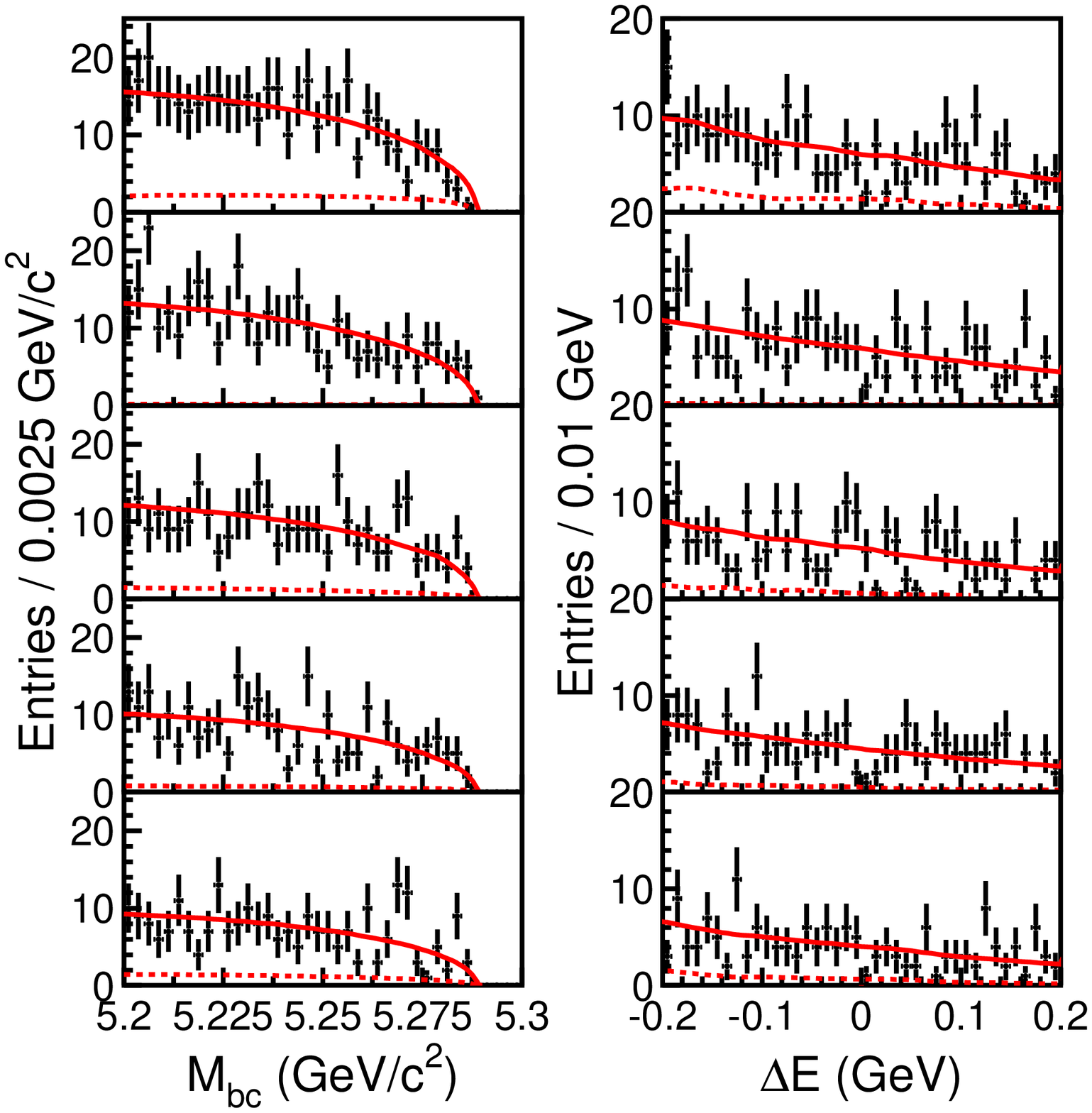}
\includegraphics[width=0.48\textwidth]{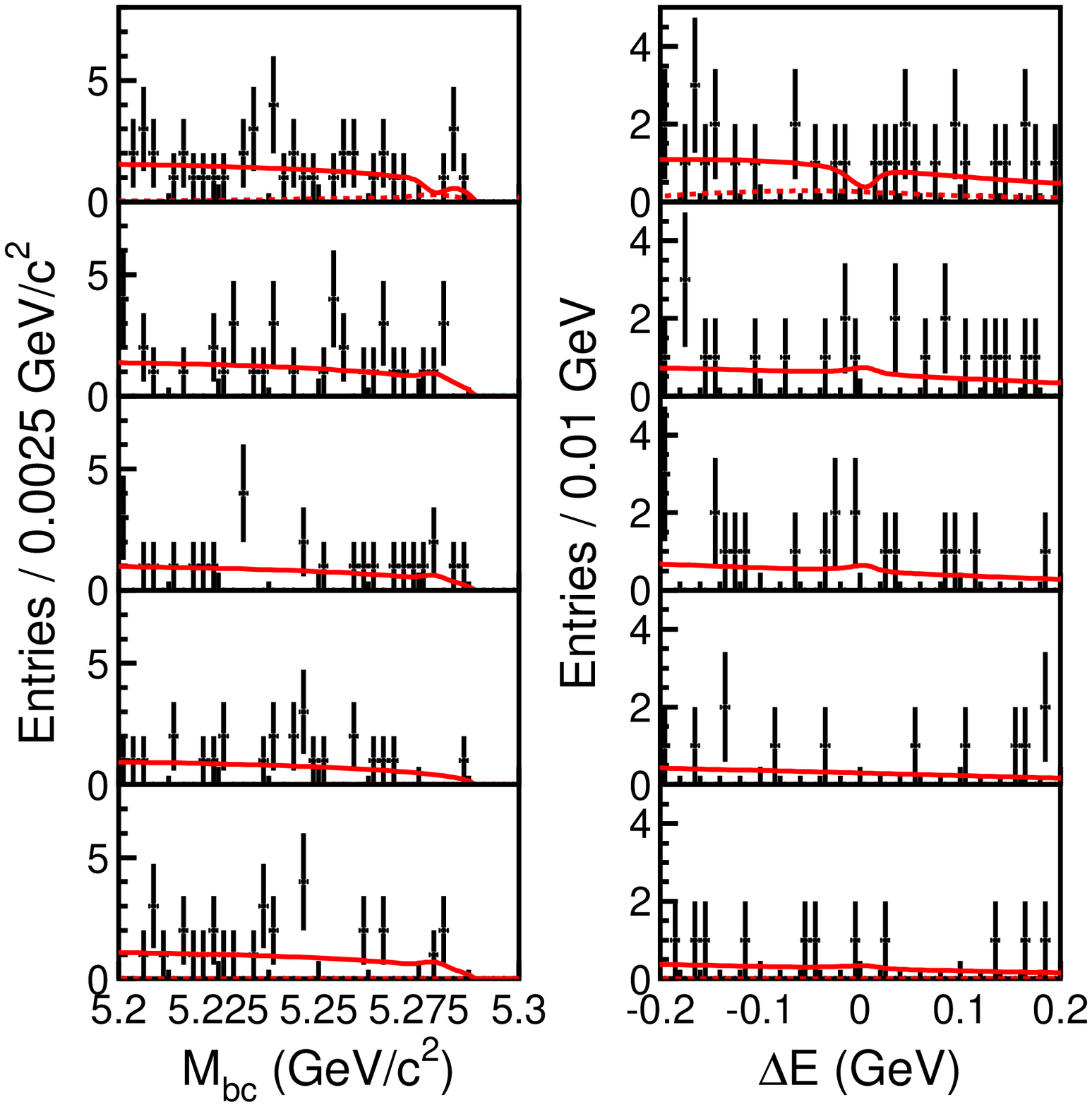}
\caption[]{The $\mb$ distributions in the $\de$ signal region and the
  $\de$ distributions in the $\mb$ signal region for the $\ekskpi$ (left) and
$\epp$ modes (right) in 100 MeV bins of $M(\eta_c\gamma)$ for $3.22 <
M(\eta_c\gamma) \le 3.72$ GeV/$c^2$. The distributions are shown in the 
increasing order of $M(\eta_c\gamma)$ from the top to the bottom. The
solid curves are the results of the fits. The dashed curves represent
background components from $B$ decays.}
\label{fig:rangeII}
\end{center}
\end{figure}

To check for possible binning effects, we determine the branching
fractions for the $  \eta_c\gamma$ invariant mass ranges that are
shifted by 50 MeV/$c^2$ with respect to the nominal range.
The results of these fits are shown in Fig. ~\ref{fig:rangeII}. 
The signal yields and the detection efficiencies are given in
Table~\ref{tab:fit}. No significant signals are observed for the range
$3.22 < M( \eta_c\gamma) \le 3.72$ GeV/$c^2$.

\begin{table}[htb]
\begin{center}
\caption[]{Detection efficiencies and signal yields in 100 MeV/$c^2$
bins of $M(\eta_c\gamma)$.}
\begin{tabular}{ccccc}
\hline
&\multicolumn{2}{c}{$\ekskpi$}&\multicolumn{2}{c}{$\epp$} \\
$M( \eta_c\gamma)$ (GeV/$c^2$) & $\varepsilon$ (\%)& Yield &
$\varepsilon$ (\%) & Yield \\
\hline
\hline
$3.17 - 3.27$  &  $4.39\pm0.10$ & $-3.2^{+5.3}_{-4.3}$ & $11.7\pm0.2$
& $-1.7^{+2.8}_{-2.1}$ \\
$3.27 - 3.37$  &  $4.62\pm0.10$ & $7.0^{+6.0}_{-5.2}$ & $12.6\pm0.2$ &
$-1.2^{+1.8}_{-1.3}$ \\
$3.37 - 3.47$  &  $5.31\pm0.10$ & $-3.6^{+3.8}_{-2.7}$ & $15.1\pm0.2$ &
$-0.30^{+2.3}_{-1.7}$ \\
$3.47 - 3.57$  &  $5.47\pm0.10$ & $-3.9^{+4.5}_{-3.5}$ & $15.1\pm0.2$ &
$-0.76^{+1.9}_{-1.1}$ \\
$3.57 - 3.67$  &  $5.95\pm0.11$ & $3.2^{+5.0}_{-4.3}$ & $16.6\pm0.2$ &
$1.4^{+2.0}_{-1.3}$ \\
\hline
$3.22 - 3.32$  &  $4.66\pm0.10$ & $-2.2^{+4.9}_{-4.0}$ & $12.2\pm0.2$
& $-3.8^{+2.2}_{-1.8}$ \\
$3.32 - 3.42$  &  $4.83\pm0.10$ & $0.70^{+5.0}_{-4.1}$ & $13.8\pm0.2$ 
&$1.4^{+2.5}_{-1.8}$ \\
$3.42 - 3.52$  &  $5.41\pm0.10$ & $2.1^{+5.2}_{-4.3}$ & $14.9\pm0.2$ 
&$1.30^{+2.3}_{-1.5}$ \\
$3.52 - 3.62$  &  $5.48\pm0.11$ & $-1.0^{+4.6}_{-3.8}$ & $16.2\pm0.2$ 
&$0^{+0.5}_{-0}$ \\
$3.62 - 3.72$  &  $6.02\pm0.11$ & $0.09^{+3.9}_{-3.0}$ & $16.9\pm0.2$ 
&$1.0^{+2.1}_{-1.4}$ \\
\hline
\hline
\end{tabular}
\label{tab:fit}
\end{center}
\end{table}

\begin{figure}[htb]
\includegraphics[width=0.70\textwidth]{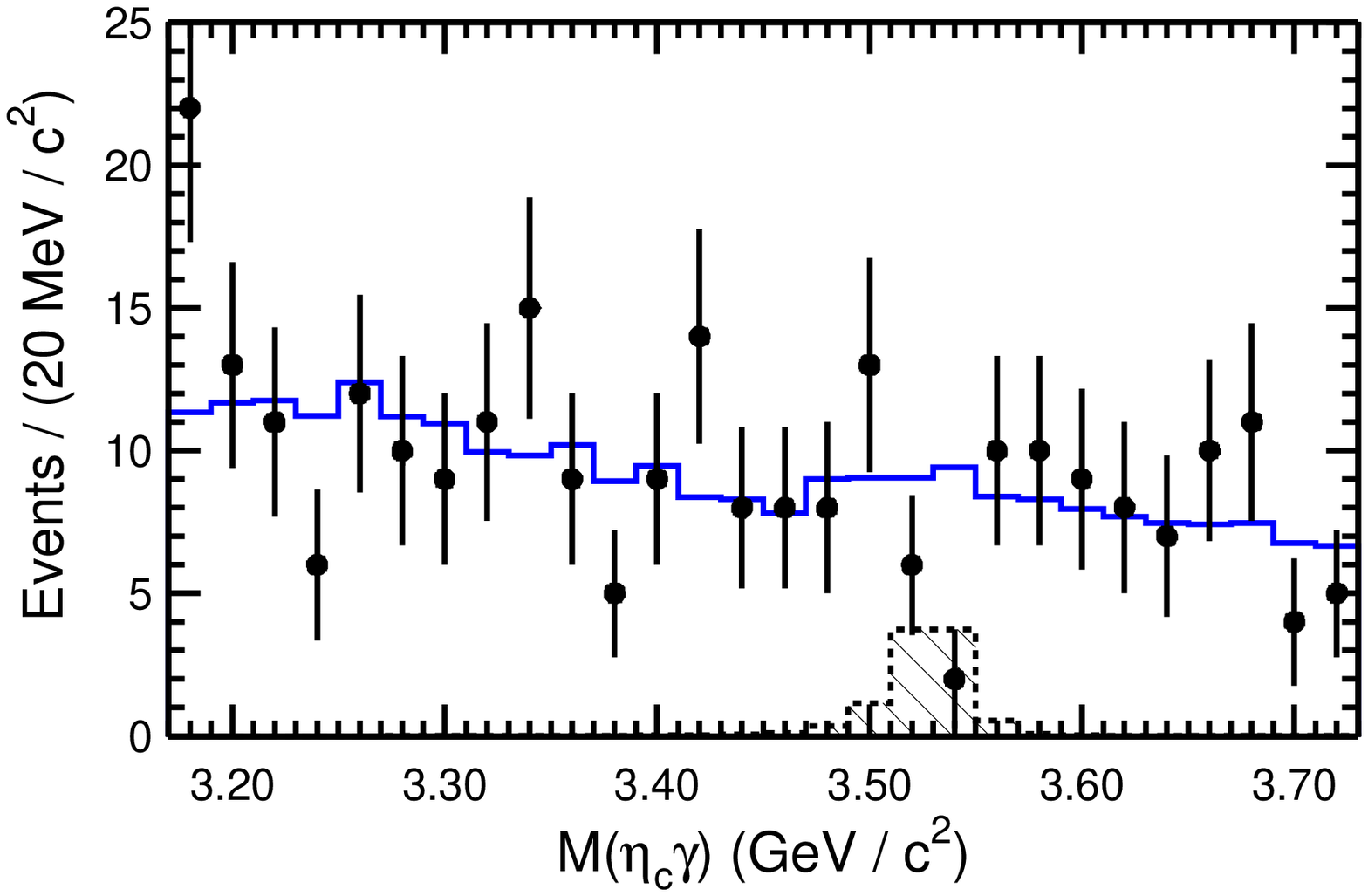}
\caption{The $M(K_S^0 K^{\pm}\pi^{\mp} \gamma)$
distributions. The points with error bars are the data in the
$M_{\rm bc}$ and $\Delta E$ signal region; the solid histogram
is the $M_{\rm bc}$ sideband data in the region $5.20<M_{\rm
bc}<5.26$ GeV/$c^2$; the hatched histogram shows the expected
contribution for a signal with a branching fraction at the observed
upper limit for a resonance with $M(\eta_c\gamma)$ = 3.527 GeV} 
\label{fig:mkkpig}
\end{figure}

Fig.~\ref{fig:mkkpig} shows the $M(\eta_c \gamma)$ distribution for
data in the $M_{\rm bc}$ and $\Delta E$ signal region (points with
error bars). The distribution is consistent with the background
determined from the $M_{\rm bc}$ sideband data in the region
$5.20<M_{\rm bc}<5.26$ GeV/$c^2$. The expected contribution for a
resonance of a mass of 3.527 GeV/$c^2$ with a branching fraction at the
observed upper limit is also shown.

To demonstrate the effectiveness of the analysis procedure 
and the method for branching fraction determination,
we examine the decay chain $B^+\to \chi_{c1} K^+$, $\chi_{c1}
\to   J/\psi\gamma$, $J/\psi \to p\bar{p}$. A clear signal of
$14.8^{+4.6}_{-3.9}$ events is observed in the $\mb$-$\de$
distribution. We use the yield and the MC detection efficiency
of 0.131 to determine the branching fraction ${\cal B}(B^+\to
\chi_{c1} K^+) = (6.1^{+1.9}_{-1.6})\times 10^{-4}$, where the error
is statistical only. This is in very good agreement with the result
calculated from the world averages~\cite{ref:pdg}, ${\cal B}(B^+\to
\chi_{c1} K^+) = (6.8 \pm 1.2)\times 10^{-4}$. 

The largest contributions to the systematic error in the detection
efficiency are the uncertainties in the efficiencies for tracking,
particle identification and photon detection. The errors in kaon, pion
and proton identification efficiencies are obtained from kinematically
selected $D^{*+}\to D^0\pi^+, D^0\to K^-\pi^+$ and $\Lambda\to p\pi^-$
decays in the data.  We apply correction factors of
0.975 and 0.941 for the pion and proton detection efficiencies,
respectively. The particle identification systematic error is 5.1\%
for the $\ekskpi$ mode and 6.6\% for the $\epp$ mode. The uncertainty
in photon detection efficiency is determined to be 5\% using a $\eta
\to \gamma \gamma$ data sample. The uncertainty in the $\eta_c$ vertex
reconstruction is estimated to be 2\% using a
$\phi \to K^{+}K^{-} $ sample. The systematic error due to the
modeling of the likelihood ratio cut is determined to be 4\% using
$B^+\to \bar{D}^0\pi^+$ events reconstructed in data. We also include
the MC statistical uncertainty and the uncertainty in the number of
$B\bar{B}$ pairs in the data sample. The sources of systematic error are
combined in quadrature to obtain the final systematic error in the
detection efficiency, which is 10.3\% for the $\ekskpi$ mode and
10.1\% for the $\epp$ mode.

The uncertainty in the signal yield from the fit is determined by
varying the mean of the signal for $\mb$ by 0.5 ${\rm MeV}/c^2$, and
all other shape parameters of the signal and the background by
1$\sigma$ of the measured errors. The results are combined in
quadrature to obtain the total uncertainty, which depends on
$M(\eta_c\gamma)$ bin and ranges from $\pm0.5$ to $\pm2.1$ for the
$\ekskpi$ mode and from $\pm0.0$ to $\pm1.3$ for the $\epp$ mode.

We combine the likelihoods for the $\ekskpi$ and $\epp$ modes, taking
into account the respective systematic errors in the detection
efficiencies and uncertainties in the signal yields. We determine upper
limits at $90\%$ confidence level (C.L.) on the branching fractions
for $B^+ \to \eta_{c}\gamma  K^+$.  The results are given in
Table~\ref{tab:ul} in bins of $M(\eta_c\gamma)$.
~\begin{table}[htb]
\begin{center}
\caption[]{Upper limits at 90\% C.L. on branching fractions for $B^+ \to
\gamma \eta_c K^+$ in bins of $M( \eta_c\gamma)$.}
\begin{tabular}{cc}
\toprule
$M( \eta_c\gamma)$ (GeV/$c^2$) & Branching Fraction \\
\hline
\hline
 $3.17 - 3.27$   & $<5.9\times10^{-5}$ \\
 $3.27 - 3.37$   & $<8.6\times10^{-5}$ \\
 $3.37 - 3.47$   & $<3.2\times10^{-5}$ \\
 $3.47 - 3.57$   & $<3.8\times10^{-5}$ \\
 $3.57 - 3.67$   & $<5.8\times10^{-5}$ \\
\hline
 $3.22 - 3.32$  &  $<4.7\times10^{-5}$ \\
 $3.32 - 3.42$  &  $<6.7\times10^{-5}$ \\
 $3.42 - 3.52$  &  $<6.2\times10^{-5}$ \\
 $3.52 - 3.62$  &  $<2.8\times10^{-5}$ \\
 $3.62 - 3.72$  &  $<3.9\times10^{-5}$ \\
\hline
\hline
\end{tabular}
\label{tab:ul}
\end{center}
\end{table}

\begin{figure}[htb]
\includegraphics[width=0.84\textwidth]{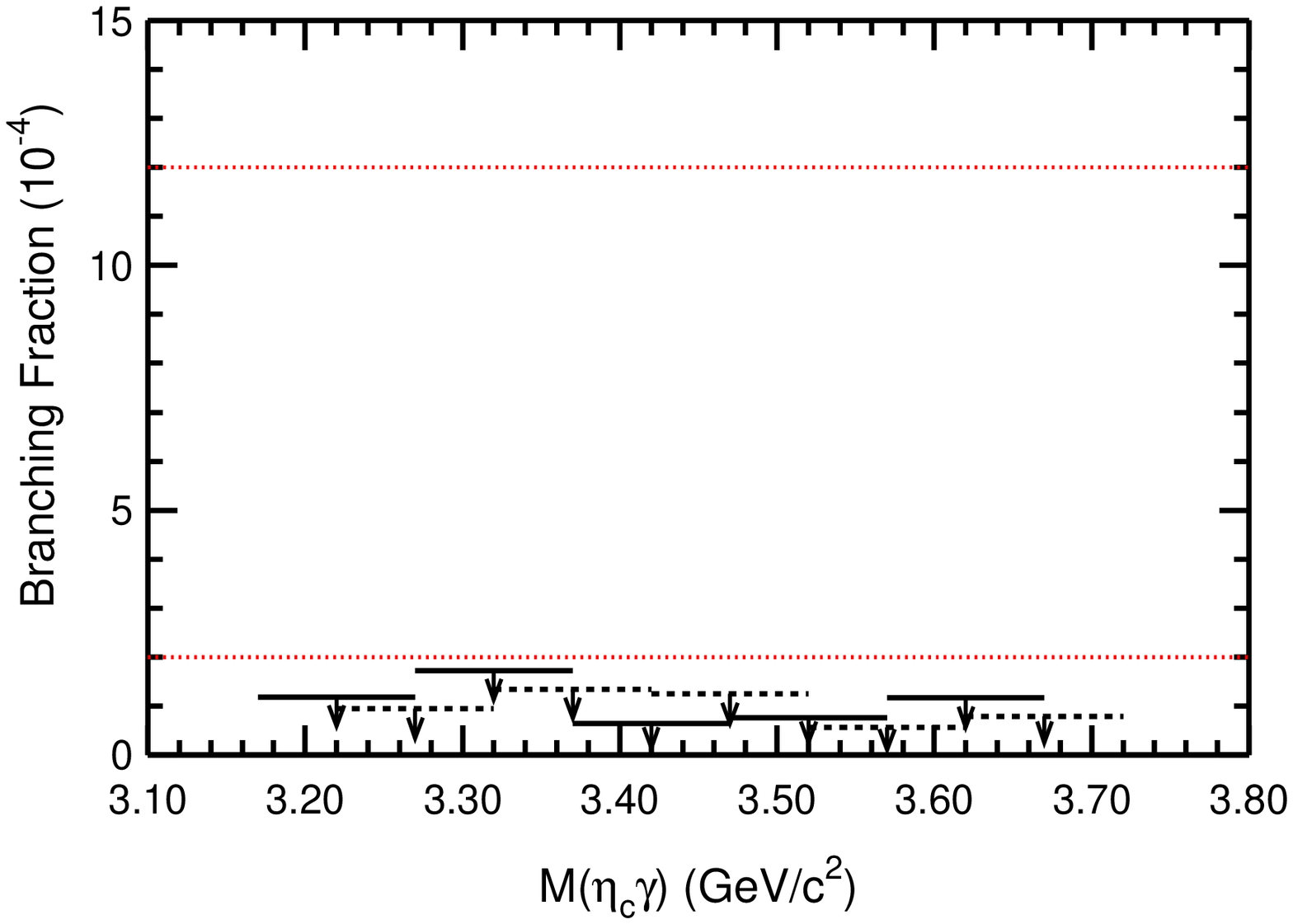}
\caption[Upper limits at 90\% C.L. on branching fractions for $B^+
\to h_c K^+$ for different assumptions of the $h_c$ mass,
assuming ${\cal B}(h_c \to  \eta_c\gamma)=0.5$. 
The dashed lines
with arrows show the upper limits for mass bins shifted by 50 MeV/c$^2$
with respect to the nominal ones.
The dotted lines represent the
theoretical range for the branching fraction obtained by Colangelo, 
Fazio and Pham]{Upper limits at 90\% C.L. on the branching fractions for
$B^+ \to h_c K^+$ as a function of the $h_c$ mass,
assuming ${\cal B}(h_c\to  \eta_c\gamma)=0.5$. 
The dashed lines
with arrows show the upper limits for mass bins shifted by 50 MeV/c$^2$
with respect to the nominal ones.
The dotted lines represent the
theoretical range for the branching fraction obtained by Colangelo,
Fazio and Pham~\cite{ref:colangelo}.} 
\label{fig:ul}
\end{figure}

In summary, 
we have searched for the $h_c$ meson in the decay chain $B^+\to h_c K^+$,
$h_c \to  \eta_c\gamma$, where the $\eta_c$ is reconstructed in the
$K_S^0 K^{\pm}\pi^{\mp}$ and $ p\bar{p}$ modes. No significant signals are
seen for $3.17<M(\gamma\eta_c)\le 3.67$ GeV/$c^2$. We obtain upper
limits on the branching fractions for $B^+ \to  \gamma \eta_c K^+$ for
different $\eta_c\gamma$ invariant mass ranges. Assuming
${\cal B}(h_c \to \gamma \eta_c)=0.5$, these results give
90\% C.L. upper limits on
branching fractions for $B^+ \to h_c K^+$ as a function of
the $h_c$ mass. The results are shown in Fig.~\ref{fig:ul}. For
$M_{h_c}=3.527$ GeV/$c^2$, we find ${\cal B}(B^+\to h_c K^+)<3.8\times
10^{-5}$. This is below
the lower bound on the $B\to h_c K$ branching 
fraction obtained by Colangelo,
Fazio and Pham~\cite{ref:colangelo}, which is
${\cal{B}}(B\to h_c K)=(2-12)\times 10^{-4}$. These results are
comparable to the upper limit for $B^+\to \chi_{c2} K^+$\cite{ref:chic2}
but below the measured rate for $B^+\to \chi_{c0} K^+$, two other
non-factorizable decays. 
The upper limits obtained in this paper assume 
${\cal B}(h_c \to \gamma \eta_c)=0.5$ and therefore must be renormalized
when this $h_c$ absolute branching fraction is measured. These results
may also be used to constrain branching fractions of other
charmonium or charmonium-like states that decay to $\eta_c \gamma$.

We thank the KEKB group for the excellent operation of the
accelerator, the KEK Cryogenics group for the efficient
operation of the solenoid, and the KEK computer group and
the National Institute of Informatics for valuable computing
and Super-SINET network support. We acknowledge support from
the Ministry of Education, Culture, Sports, Science, and
Technology of Japan and the Japan Society for the Promotion
of Science; the Australian Research Council and the
Australian Department of Education, Science and Training;
the National Science Foundation of China under contract
No.~10175071; the Department of Science and Technology of
India; the BK21 program of the Ministry of Education of
Korea and the CHEP SRC program of the Korea Science and
Engineering Foundation; the Polish State Committee for
Scientific Research under contract No.~2P03B 01324; the
Ministry of Science and Technology of the Russian
Federation; the Ministry of Education, Science and Sport of
the Republic of Slovenia; the National Science Council and
the Ministry of Education of Taiwan; and the U.S.\
Department of Energy.


%


\begin{thebibliography}{99}
\bibitem{ref:cc} The charge conjugate decay mode is implied throughout
  this paper unless otherwise stated.
\end{thebibliography}
\end{document}